\documentstyle[preprint,aps]{revtex}
\begin{document}
\title{On-Line AdaTron Learning of Unlearnable Rules}
\date{Received:\mbox{\hspace{1.5in}}}
\author{Jun-ichi Inoue and Hidetoshi Nishimori}
\address{Department of Physics, Tokyo Institute of 
Technology, \\ Oh-okayama, Meguro-ku, Tokyo 152, Japan}
\maketitle
\begin{abstract}
We study the on-line AdaTron learning of linearly 
non-separable rules by a simple perceptron.
Training examples are provided by a perceptron
with a non-monotonic transfer function which reduces
to the usual monotonic relation in a certain limit.
We find that, although the on-line 
AdaTron learning is a powerful algorithm for the 
learnable rule, it does not give the best possible
generalization error for unlearnable problems.
Optimization of the learning rate is shown to
greatly improve the performance of the AdaTron
algorithm, leading to the best possible generalization
error for a wide range of the parameter which controls
the shape of the transfer function.
\end{abstract}
\section{INTRODUCTION}
The problem of learning is one of the most interesting aspects
of feed-forward neural networks
\cite{Hertz91,Watkin93,Opper95}.
Recent activities in the theory of learning have gradually
shifted toward the issue of on-line learning.
In the on-line learning scenario, 
the student is trained only by 
the most recent example which is never referred to  again.
In contrast, in the off-line (or batch) 
learning scheme, the student is given 
a set of examples repeatedly and memorizes
these examples so as to minimize the global cost function.
Therefore, the on-line learning 
has several advantages over the off-line method. 
For example, it is not necessary for the student 
to memorize the whole set of examples, which saves
a lot of memory space.
In addition, theoretical analysis of on-line learning
is usually much less complicated than that of
off-line learning which often makes use of the
replica method.

In many of the studies of learning, authors assume
that the teacher and student networks have the same
structures. The problem is called learnable in these cases .
However, in the real world we find innumerable unlearnable
problems where the student is not able to perfectly
reproduce the output of teacher in principle.
It is therefore both important and interesting to devote
our efforts to the study of learning unlearnable rules.

If the teacher and student have the same structure, 
a natural strategy of learning is to modify
the weight vector of student 
${\bf J}$ so that this approaches
teacher's weight ${\bf J}^{0}$ as 
quickly as possible. 
However, if the teacher and student have different structures, 
the student trained to satisfy ${\bf J}={\bf J}^{0}$ 
sometimes cannot generalize the unlearnable rule better than 
the student with
${\bf J}{\neq}{\bf J}^{0}$. 
Several years ago, Watkin and Rau \cite{Watkin92} 
investigated the off-line learning of unlearnable rule 
where the teacher is a perceptron with a non-monotonic 
transfer function while the student is a simple perceptron. 
They discussed the case where the number of examples
is of order unity and therefore did not derive the asymptotic
form of the generalization error in the limit
of large number of training examples.
Furthermore, as they used the replica method 
under the replica symmetric ansatz,
the result may be unstable against replica symmetry breaking.

For such a type of non-monotonic transfer function, a lot of 
interesting phenomena have been reported.
For example, the critical loading rate of the model
of Hopfield type \cite{Morita90,Nishi93,Ino96} or the optimal 
storage capacity of perceptron \cite{Boff93} is known to
increase dramatically by non-monotonicity. 
It is also worth noting that perceptrons with the non-monotonic 
transfer function can be regarded as a toy model of a
multilayer perceptron, a parity machine \cite{Monasson94}.

In this context, Inoue, Nishimori and Kabashima \cite{Inoue96}
recently investigated the problem of
on-line learning of unlearnable rules where the teacher 
is a non-monotonic perceptron: the output of the teacher is
$T_{a}(v)={\rm sign}[v(a-v)(a+v)]$,
where $v$ is the input potential of the teacher
 $v\,{\equiv}\,
\sqrt{N}({\bf J}^{0}{\cdot}{\bf x})$, 
with ${\bf x}$ being a training example, and the student is a simple perceptron. 
For this system, difficulties of 
learning for the student can be controlled by 
the width $a$ of the reversed wedge. 
If $a=\infty$ or $a=0$, the student can learn 
the rule perfectly and the generalization error 
decays to zero as ${\alpha}^{-1/3}$ for the conventional 
perceptron learning algorithm
and ${\alpha}^{-1/2}$ for the Hebbian learning algorithm,
where $\alpha$ is the number of presented examples, $p$,
divided by the number of input nodes, $N$. 
For finite $a$, the student cannot generalize perfectly and 
the generalization error converges exponentially
to a non-vanishing $a$-dependent value.

In this paper we investigate the generalization ability 
of student trained by the on-line AdaTron learning algorithm 
with examples generated by the above-mentioned
non-monotonic rule. 
The AdaTron learning is a powerful method for learnable
rules both in on-line and off-line modes in the sense
that this algorithm gives a fast decay,
proportional to $\alpha^{-1}$,
of the generalization error \cite{Biehl94,Anlauf89,Riegler95},
in contrast to the $\alpha^{-1/3}$ and $\alpha^{-1/2}$ decays
of the perceptron and Hebbian algorithms.
We investigate the performance of the AdaTron learning
algorithm in the unlearnable situation and discuss the
asymptotic behavior of the generalization error.

This paper is organized as follows.
In the next section, we explain the generic 
properties of the generalization error for our system and 
formulate the on-line AdaTron learning.
Some of the results of our previous paper \cite{Inoue96}
are collected here concerning the perceptron and Hebbian
learning algorithms which are to be compared with
the AdaTron learning.
Section III deals with the conventional AdaTron learning
both for learnable and unlearnable rules.
In Sec. IV we investigate the effect of optimization 
of the learning rate.
In Sec. V the issue of optimization is
treated from a different point of view where we do
not use the parameter $a$, which is unknown to the student,
in the learning rate. 
In last section we summarize our 
results and discuss several future problems. 
\section{THE MODEL SYSTEM}
Let us first fix the notation.  The input signal comes from $N$
input nodes and is represented by an $N$-dimensional
vector $\bf x$.
The components of $\bf x$ are randomly drawn from a
uniform distribution and then $\bf x$ is normalized to unity.
Synaptic connections from input nodes to the student perceptron
are also expressed by an $N$-dimensional vector $\bf J$
which is not normalized.
The teacher receives the same input signal $\bf x$ through the
normalized synaptic weight vector ${\bf J}^0$.
The generalization error is ${\epsilon}_{\rm g}\,{\equiv}\,
{\ll}{\Theta}(-T_{a}(v)S(u)){\gg}$, where 
$S(u)={\rm sign}(u)$ is the student output with the
internal potential $u\,{\equiv}\,
\sqrt{N}({\bf J}{\cdot}{\bf x})/|{\bf J}|$ and 
${\ll}{\cdots}{\gg}$ stands for the average over the distribution
function
\begin{equation}
P_{R}(u,v)=\frac{1}{2{\pi}\sqrt{1-R^{2}}}
\,{\exp}\left[
-\frac{(u^{2}+v^{2}-2Ruv)}{2(1-R^{2})}
\right] .
\label{dist1}
\end{equation}
Here $R$ stands for the overlap between the
teacher and student weight vectors,
$R\,{\equiv}\,({\bf J}^{0}{\cdot}{\bf J})/|{\bf J}^{0}||{\bf J}|$.
This distribution has been 
derived from randomness of ${\bf x}$ and 
is valid in the limit $N{\rightarrow}\infty$. 

The generalization error ${\epsilon}_{\rm g}$ 
is easily calculated as a function of $R$ as follows \cite{Inoue96} 
\begin{equation}
\epsilon_{\rm g}=E(R)\equiv 2\int_{a}^{\infty}Dv\, H\left(
-\frac{Rv}{\sqrt{1-R^{2}}}
\right)
+2\int_{0}^{a}\, Dv\,H\left(
\frac{Rv}{\sqrt{1-R^{2}}}
\right) ,
\label{ge}
\end{equation}
where $H(x)\,{\equiv}\,\int_{x}^{\infty}Dt$ 
with $Dt\,{\equiv}\,{\exp}(-t^{2}/2)/\sqrt{2\pi}$.
It is important that this expression is independent of 
specific learning algorithm.
Minimization of $E(R)$ with respect to $R$ gives the 
theoretical lower bound, or the best possible 
value,  of the generalization error for given $a$. 
In Fig. 1 we show $E(R)$ for several values of $a$. 
This figure indicates that the generalization error goes to zero 
if the student is trained so that the overlap
$R$ becomes 1 for $a=\infty$ and $R=-1$ for $a=0$. 
If the parameter $a$ is larger than some critical
value $a_{\rm c1}=\sqrt{2{\log}2}=1.177$,
$E(R)$ decreases monotonically 
from $1$ to $0$ as $R$ increases from $-1$ to $1$. 
When $a$ is smaller than $a_{\rm c1}$, a local minimum appears at  
$R=R_{*}\equiv -\sqrt{(2{\log}2-a^{2})/2{\log}2}$,
but the global minimum is still at $R=1$
as long as $a$ is larger than $a_{\rm c2}=0.80$. 
If $a$ is less than $a_{\rm c2}$, the
global minimum is found at $R=R_{*}$, not at $R=1$.
This situation is depicted in Figs. 2 and 3 
where we show the optimal overlap $R$
giving the smallest value of $E(R)$ and the corresponding
best possible value of the generalization error as functions of $a$. 
From these two figures, we see that the
optimal overlap which gives the theoretical 
lower bound shows a first-order phase 
transition at $a=a_{\rm c2}$. 

Therefore, our efforts should be directed to finding 
the best strategy which gives the best possible value of 
the generalization error for a wide range of the parameter $a$. 

It may be useful to review some of the results of, 
Inoue, Nishimori and Kabashima \cite{Inoue96} 
who studied the present problem under the perceptron
and Hebbian algorithms.
For the conventional perceptron learning, the
generalization error decays to zero as 
${\alpha}^{-1/3}$ if the rule is learnable ($a=\infty$), 
whereas it converges to a non-vanishing value
$E(R=1-2{\Delta})$, where ${\Delta}\,{\equiv}\,
{\exp}(-a^{2}/2)$,  exponentially for the unlearnable case. 
This value of $E(R)$ is larger than the best possible value
as seen in Fig. 3.
Introduction of optimization processes of the learning
rate improves the performance significantly
in the sense that the generalization error then
converges to the best possible value when $a>a_{\rm c2}$.
For the conventional Hebbian learning, the
generalization error decays to the theoretical lower bound
as ${\alpha}^{-1/2}$ not only in the learnable limit 
$a{\rightarrow}\infty$ but for a finite range of $a$,
$a>a_{\rm c1}$. 
However, for $a<a_{\rm c1}$, the generalization error 
does not converge to the optimal value. 
\section{LEARNING DYNAMICS}
The on-line training dynamics of the AdaTron algorithm is
\begin{equation}
{\bf J}^{m+1}={\bf J}^{m} 
-g(\alpha)\,u\, {\Theta}(-T_{a}(v)S(u))\,
{\bf x},
\label{dyn1}
\end{equation}
where $m$ stands for the number of presented patterns
and $g(\alpha)$ is the leaning rate.
It is straightforward to obtain the recursion equations for 
the overlap $R^{m}=({\bf J}^{m}{\cdot}{\bf J}^{0})/|{\bf J}^{m}||{\bf J}^{0}|$ and
the length of the student weight vector
$l^{m}=|{\bf J}^{m}|/\sqrt{N}$. 
In the limit $N{\rightarrow}\infty$, 
these two dynamical quantities become self-averaging 
with respect to the random training data $\bf x$.
For continuous time ${\alpha}=m/N$ 
in the limit $N{\rightarrow}\infty$, $m{\rightarrow}\infty$ 
with $\alpha$ kept finite, the
evolutions of $R$ and $l$ are given by the following
differential equations \cite{Inoue96}:
\begin{equation}
\frac{dl}{d\alpha}=\frac{g^2 E_{\rm Ad}}{2l}
   -gE_{\rm Ad}
\label{dlda}
\end{equation}
\begin{equation}
\frac{d R}{d \alpha}=
-\frac{Rg^2 E_{\rm Ad}}{2l^2}
+\frac{gE_{\rm Ad}R-G_{\rm Ad}}{l},
\label{drda}
\end{equation}
where 
\begin{eqnarray}
E_{\rm Ad}\,{\equiv}\,{\ll}u^{2}{\Theta}(-T_{a}(v)S(u)){\gg}=
\sqrt{\frac{2}{\pi}}\int_{0}^{\infty}
u^{2}Du\,H_{a}(u,R)
\label{ead}
\end{eqnarray}
with
\begin{equation}
H_{a}(u,R)\,{\equiv}\,
H\left(
\frac{a-Ru}{\sqrt{1-R^{2}}}
\right)
+
H\left(
\frac{Ru}{\sqrt{1-R^{2}}}
\right)
-
H\left(
\frac{a+Ru}{\sqrt{1-R^{2}}}
\right)
\label{ha}
\end{equation}
and
\begin{eqnarray}
G_{\rm Ad}\,{\equiv}\,{\ll}uvT_{a}(v){\Theta}(-T_{a}(v)S(u)){\gg} 
    \hspace{2.0in}\nonumber \\
\mbox{}=\frac{1}{\pi}(1-R^{2})^{3/2}
\left[
2\, {\exp}\left(
-\frac{a^{2}}{2(1-R^{2})}
\right) 
-1
\right]\hspace{1.0in}\nonumber \\
\mbox{}\hspace{.4in}+\sqrt{\frac{2}{\pi}}\, Ra
(\sqrt{1-R^{2}}){\Delta}
\left[
1-2H\left( \frac{Ra}{\sqrt{1-R^{2}}}\right)
\right]
+RE_{\rm Ad}.
\label{gg}
\end{eqnarray}
Equations (\ref{dlda}) and (\ref{drda}) determine the learning
process.  In the rest of the present section we restrict ourselves
to the case of $g=1$ corresponding to the conventional
AdaTron learning.
\subsection{Learnable case}
We first consider the case of $g(\alpha)=1$ and 
$a=\infty$, the learnable rule.
We investigate the asymptotic behavior of the generalization
error when $R$ approaches 1,
$R=1-{\varepsilon}$, 
${\varepsilon}{\rightarrow}0$ and $l=l_{0}$, a constant.
From Eqs. (\ref{ead}) and (\ref{gg}), we find 
$E_{\rm Ad}\,{\sim}\,c\,{\varepsilon}^{3/2}$ and 
$G_{\rm Ad}\,{\sim}\,(c-2\sqrt{2}/\pi)\,{\varepsilon}^{3/2}$ 
with $c=8/(3\sqrt{2}\pi)$.
Then Eq. (\ref{drda}) is solved as
${\varepsilon}=(2/k)^{2}{\alpha}^{-2}$ with 
\begin{equation}
k\,{\equiv}\,\frac{2l_{0}-1}{2l_{0}^{2}}c
+\frac{2\sqrt{2}-c\pi}
{{\pi}l_{0}}. 
\label{kk}
\end{equation}
Using this equation and Eq. (\ref{ge}), we obtain
the asymptotic form of the generalization error as
\begin{equation}
{\epsilon}_{\rm g}=E(R)\sim \frac{\sqrt{2\varepsilon}}{\pi}
=\frac{2\sqrt{2}}{{\pi}k}\frac{1}{\alpha}.
\label{eg2}
\end{equation}
The above expression of the generalization error depends
on $l_0$, the asymptotic value of $l$, through $k$.
Apparently $l_0$ is a function of the initial value of $l$
as shown in Fig. 4.
A special case is $l_0=1/2$ in which case $l$ does not
change as learning proceeds as is apparent from
Eq. (\ref{dlda}) as well as from Fig. 4.  Such a
constant-$l$ problem was studied by
Biehl and Riegler \cite{Biehl94} who concluded
\begin{equation}
{\epsilon}_{\rm g}=\frac{3}{2\alpha}
\label{ge3}
\end{equation}
for the AdaTron algorithm.
Our formula (\ref{eg2}) reproduces this result when $l_0=1/2$.
If one takes $l_0$ as an adjustable parameter, it is possible
to minimize $\epsilon_{\rm g}$ by maximizing $k$ in the
denominator of Eq. (\ref{eg2}).  The smallest value
of $\epsilon_{\rm g}$ is achieved when $l_0=\pi c/2\sqrt{2}$,
yielding
\begin{equation}
{\epsilon}_{\rm g}=\frac{4}{3\alpha}
\label{ge4}
\end{equation}
which is smaller than Eq. (\ref{ge3}) for a fixed $l$.
We therefore have found that the asymptotic behavior
of the generalization error depends upon whether or
not the student weight vector is normalized 
and that a better result is obtained for the un-normalized case.
We plot the generalization error for the present learnable case
with the initial value of $l_{\rm init}=0.1$ in Fig. 5. 
We see that the Hebbian learning 
has the highest generalization ability 
and the AdaTron learning shows the slowest 
decay among the three algorithms 
in the initial stage of learning. 
However, as the number of presented patterns increases,
the AdaTron algorithm eventually achieves the smallest
value of the generalization error.
In this sense the AdaTron learning algorithm is the
most efficient learning strategy among the three
in the case of the learnable rule.
\subsection{Unlearnable case}
For unlearnable case, there can exist only 
one fixed point $l_{0}=1/2$. 
This reason is, for finite $a$, $E_{\rm Ad}$ appearing in Eq. (\ref{dlda}) 
does not vanish 
in the limit of large ${\alpha}$ and 
$E_{\rm Ad}$ has a finite value for $a{\neq}\infty$. 
For this finite $E_{\rm Ad}$, 
the above differential equation 
has only one fixed point $l_{0}=1/2$. 
In contrast, for the learnable case, 
$E_{\rm Ad}$ behaves as $E_{\rm Ad}\,{\sim}\,c\,{\varepsilon}^{3/2}$ 
in the limit of ${\alpha}{\rightarrow}\infty$ 
and thus $dl/d\alpha$ becomes zero irrespective of $l$ 
asymptotically. 
We plot trajectories in the $R$-$l$ plane for $a=2$ 
in Fig. 6 and the corresponding generalization error 
is plotted in Fig. 7 as an example. 
From Fig. 6, we see that 
the destination of $l$ is $1/2$ for 
all initial conditions. 
Figure 7 tells us that for the unlearnable case $a=2$, 
the AdaTron learning has the lowest generalization ability 
among the three. 
We should notice that the generalization error decays 
to 
its asymptotic value, the residual error ${\epsilon}_{\rm min}$,  
as ${\epsilon}_{\rm g}-{\epsilon}_{\rm min}\,{\sim}\,{\alpha}^{-1/2}$
for the Hebbian learning and 
decays exponentially for perceptron learning \cite{Inoue96}. 
The residual error of the Hebbian learning 
${\epsilon}_{\rm min}=2H(a)$ is also the best  
possible value of the generalization error for $a>a_{\rm c{2}}$ 
as seen in Fig. 3.
In Fig. 8 we also plot the generalization error 
of the AdaTron algorithm 
for several values of $a$.
For the AdaTron learning of 
the unlearnable case, the generalization error converges to 
a non-optimal value $E(R_{0})$ 
exponentially.

For all unlearnable cases, the $R$-$l$ flow is attracted into 
the fixed point $(R_{0},1/2)$, where $R_{0}$ is obtained from
\begin{equation}
\frac{d R}{d \alpha}\,{\Biggm |}_{l=\frac{1}{2},\,R=R_{0}}=
-2G_{\rm Ad}(R_{0})=0. 
\label{rfix}
\end{equation}
The solution $R_{0}$ of the 
above equation is not the optimal value because 
the optimal value of the present  
learning system is $R_{\rm opt}=1$ for $a>a_{\rm c2}$ and 
$R_{\rm opt}=R_{*}=-\sqrt{(2{\log}2-a^{2})/2{\log}2}$ 
for $a<a_{\rm c2}$ \cite{Inoue96}. 

From Figs. 3 and 7, we see that 
the residual error ${\epsilon}_{\rm min}$ 
of the AdaTron learning is larger than that of 
the conventional perceptron learning. 
Therefore, we conclude that 
if the student learns from the 
unlearnable rules, 
the on-line AdaTron algorithm becomes the worst strategy among three 
learning algorithms as we discussed above 
although for the learnable case,  the on-line 
AdaTron learning 
is a sophisticated algorithm and 
the generalization error decays to zero as 
quickly as the off-line learning \cite{Opper90}. 
\section{OPTIMIZATION}
In the previous section, we saw that the on-line AdaTron learning 
fails to get the best possible value of 
the generalization error for the unlearnable case and 
its residual error ${\epsilon}_{\rm min}$ is 
larger than that  of the conventional perceptron 
learning or Hebbian learning. 
We show that it is possible to overcome this difficulty. 

We now consider an optimization the learning rate 
$g(\alpha)$ \cite{Inoue96}.
This optimization procedure 
is different from the technique of 
Kinouchi and Caticha \cite{Kinouchi92}.
As the optimal value  of $R$ which gives the 
best possible value of the generalization error 
is $R_{\rm opt}=1$ for 
$a>a_{\rm c2}$, we determine $g(\alpha)$ 
so that $R$ is accelerated to become $1$. 
In order to determine $g$ using the above strategy, we maximize 
the right hand side 
of Eq. (\ref{drda}) 
with respect to $g(\alpha)$ and obtain 
$g_{\rm opt}=(E_{\rm Ad}R-G_{\rm Ad})/RE_{\rm Ad}$. 
Using this optimal learning rate, Eqs. 
(\ref{dlda}) and (\ref{drda}) are rewritten as follows 
\begin{equation}
\frac{d l}{d \alpha}=
-\frac{(E_{\rm Ad}R-G_{\rm Ad})(E_{\rm Ad}R+G_{\rm Ad})}
{2R^{2}E_{\rm Ad}}l
\label{dlda2}
\end{equation}
\begin{equation}
\frac{d R}{d \alpha}=\frac{(E_{\rm Ad}R-G_{\rm Ad})^{2}}
{2RE_{\rm Ad}}.
\label{drda2}
\end{equation}

For the learnable case,  
we obtain the asymptotic form 
of the generalization error 
from Eqs. (\ref{dlda2}) and (\ref{drda2}) by the same relation 
$R=1-{\varepsilon}$, ${\varepsilon}{\rightarrow}0$
as we used for the case of $g=1$ as 
\begin{equation}
{\epsilon}_{\rm g}=\frac{4}{3\alpha}.
\label{ge5}
\end{equation}
This is the same asymptotic behavior as that obtained by 
optimizing the initial value  of $l$ 
as we saw in the previous section. 

Next we investigate the unlearnable case. 
The asymptotic forms of $E_{\rm Ad}$ and 
$E_{\rm Ad}R-G_{\rm Ad}$ in the limit of 
${\alpha}{\rightarrow}\infty$ are obtained as
\begin{equation}
E_{\rm Ad}\,{\sim}\,2H(a)+\sqrt{\frac{2}{\pi}}a{\Delta}
\label{ase}
\end{equation}
and
\begin{equation}
E_{\rm Ad}R-G_{\rm Ad}\,{\sim}\,
-\frac{4a{\varepsilon}{\Delta}}{\sqrt{2\pi}}.
\label{aseg}
\end{equation}
Then we get the asymptotic solution of Eq. (\ref{drda2}) 
with respect to ${\varepsilon}$, $R=1-{\varepsilon}$,  as 
\begin{equation}
{\varepsilon}=\frac{2{\pi}H(a)+\sqrt{2\pi}\,a\,{\Delta}}
{4a^{2}{\Delta}}
\frac{1}{\alpha}.
\label{asep}
\end{equation}
As the asymptotic behavior of $E(R)$ is obtained as 
$E(R)={\epsilon}_{\rm g}=2H(a)+\sqrt{2\varepsilon}/\pi$ \cite{Inoue96}, 
we find the generalization error in the 
limit of ${\alpha}{\rightarrow}\infty$ as follows 
\begin{equation}
{\epsilon}_{\rm g}=2H(a)+\frac{\sqrt{2}}{\pi}
\sqrt{
\frac{2{\pi}H(a)+\sqrt{2\pi}a{\Delta}}
{4a^{2}{\Delta}}}
\frac{1}{\sqrt{\alpha}},
\label{gef}
\end{equation}
where $2H(a)$ is the best possible  
value of the generalization error 
for $a>a_{\rm c2}$. 
Therefore, our strategy to optimize the learning rate 
succeeds in training the 
student to obtain the optimal overlap 
$R=1$  for $a > a_{\rm c2}$.

For the perceptron learning, 
this type of optimization failed to reach 
the theoretical lower bound of the generalization error 
for $a$ exactly at $a=a_{\rm c1}=\sqrt{2{\log}2}$ 
in which case the generalization error 
is ${\epsilon}_{\rm g}=1/2$, equivalent 
to a random guess 
because for $a=a_{\rm c1}$ optimal learning 
rate vanishes \cite{Inoue96}.
In contrast, for the AdaTron learning, the optimal 
learning rate has a non-zero value even at  $a=a_{\rm c1}$. 
In this sense, the on-line AdaTron learning with optimal 
learning rate 
is superior to the perceptron learning. 
\section{PARAMETER-FREE OPTIMIZATION}
In the previous section, we were able to  
get the theoretical lower bound 
of the generalization error 
for $a>a_{\rm c{2}}$ by introducing the optimal 
learning rate $g_{\rm opt}$.
However, as the optimal learning rate 
$g_{\rm opt}$ contains a parameter $a$ unknown to the 
student, 
the above result can be regarded only as a lower 
bound of the generalization error. 
The reason is that the student 
can get information only about 
teacher's output and  no knowledge of  
$a$ or $v=\sqrt{N}({\bf J}^{0}
{\cdot}{\bf x})/|{\bf J}^{0}|$. 
In realistic situations, 
the student does not know $a$ or $v$ and 
therefore  has a larger value of 
the generalization error. 
In this section, we construct a learning 
algorithm without the unknown parameter $a$ 
using the asymptotic form of 
the optimal learning rate. 
\subsection{Learnable case}
For the learnable case,  the optimal 
learning rate is estimated in the limit 
of ${\alpha}{\rightarrow}\infty$ as 
\begin{equation}
g_{\rm opt}=\frac{E_{\rm Ad}R-G_{\rm Ad}}
{RE_{\rm Ad}}l\,{\simeq}\,
\frac{3}{2}l.
\label{asl}
\end{equation}
This asymptotic form of the 
optimal learning rate depends on ${\alpha}$ only 
through the length $l$ of student's weight vector. 
We therefore adopt $g(\alpha)$ 
proportional to 
$l$, $g(\alpha)={\eta}\,l$, also 
in the case of the parameter-free optimization 
and adjust 
the parameter $\eta$ so that the student 
obtains the best generalization ability.  
Substituting this expression into 
the differential equation (\ref{drda})
for $R$ and using 
$R=1-{\varepsilon}$ with ${\varepsilon}{\rightarrow}0$, we get 
\begin{equation}
\frac{d \varepsilon}{d \alpha}
=-F(\eta)\,{\epsilon}^{3/2}.
\label{opdrda}
\end{equation}
where we have set 
\begin{equation}
F(\eta)\,{\equiv}\,\frac{2\sqrt{2}}{\pi}{\eta}-\frac{4}{3\sqrt{2}\pi}
{\eta}^{2}.
\label{feta}
\end{equation}
This leads to ${\varepsilon}=(F(\eta)/2)^{-2}{\alpha}^{-2}$. 
Then, the generalization error is obtained from 
${\epsilon}_{\rm g}=\sqrt{2\varepsilon}/\pi$ as 
\begin{equation}
{\epsilon}_{\rm g}=\frac{2\sqrt{2}}{\pi{F}(\eta)}\frac{1}{\alpha}. 
\label{gge}
\end{equation}
In order to minimize  ${\epsilon}_{\rm g}$, 
we maximize $F(\eta)$ with respect to $\eta$.  
The optimal choice of $\eta$ in this 
sense is ${\eta}_{\rm opt}=3/2$ and we find in such a case 
\begin{equation}
{\epsilon}_{\rm g}=\frac{4}{3\alpha}.
\end{equation}
This is the same asymptotic form as the previous $a$-dependent result (\ref{ge5}). 
\subsection{Unlearnable case}
Next we consider the unlearnable case. 
The asymptotic form of the learning rate we derived 
in the previous section for 
the unlearnable case 
is 
\begin{equation}
g_{\rm opt}=\frac{E_{\rm Ad}R-G_{\rm Ad}}{RE_{\rm Ad}}
\,{\simeq}\,
-\frac{4a{\varepsilon}\Delta/\sqrt{2\pi}}
{2H(a)+\sqrt{2/\pi}a{\Delta}}l=
{\eta}\,\frac{l}{\alpha},
\label{asg2}
\end{equation}
where we used Eq. (\ref{asep}) to obtain the 
right-most equality and we set the $a$-dependent prefactor 
of $l$ as $\eta$.
Using 
this learning rate (\ref{asg2}) 
and the asymptotic forms of $E_{\rm Ad}(R=1-{\varepsilon}, 
{\varepsilon}{\rightarrow}0)$ 
and $G_{\rm Ad}(R=1-{\varepsilon}, {\varepsilon}{\rightarrow}0)$ as 
$E_{\rm Ad}\,{\sim}\,2H(a)+\sqrt{2/\pi}a{\Delta}$ and 
$G_{\rm Ad}\,{\sim}\,4a{\Delta}{\varepsilon}/\sqrt{2\pi}+E_{\rm Ad}$ 
in the limit of ${\alpha}{\rightarrow}\infty$,  
we obtain the differential equation with respect to 
${\varepsilon}$ from Eq. (\ref{drda}) as follows 
\begin{equation}
\frac{d \varepsilon}{d \alpha}=\frac{1}{2}
\left[
2H(a)+\sqrt{\frac{2}{\pi}}a{\Delta}
\right]
\frac{{\eta}^{2}}{{\alpha}^{2}}
-{\eta}
\frac{4a}{\sqrt{2\pi}}
{\Delta}\frac{\varepsilon}{\alpha}.
\label{difeq}
\end{equation}
This differential equation can be solved analytically as 
\begin{equation}
{\varepsilon}=
\frac{{\eta}^{2}
\left(
2H(a)+\sqrt{2/\pi}a{\Delta}
\right)
}
{2\left(
4a{\Delta}{\eta}/\sqrt{2\pi}-1
\right)
}
\frac{1}{\alpha}
+A\left(
\frac{\eta}{\alpha}
\right)^{4a{\Delta}{\eta}/\sqrt{2\pi}},
\label{sol}
\end{equation}
where $A$ is a constant determined by the initial condition.
Therefore, if we choose ${\eta}$ to 
satisfy $4a{\Delta}{\eta}/\sqrt{2\pi}-1>0$, 
the generalization error converges to the 
optimal value $2H(a)$ as 
\begin{eqnarray}
{\epsilon}_{\rm g}=
2H(a)+\frac{\sqrt{2\varepsilon}}{\pi} \hspace{1.6in}\nonumber \\
\mbox{}=2H(a)+\frac{\eta}{\pi}
\sqrt{\frac{
2H(a)+\sqrt{2/\pi}a{\Delta}
}
{
4a{\Delta}{\eta}/\sqrt{2\pi}-1
}
}
\frac{1}{\sqrt{\alpha}}.
\label{type1}
\end{eqnarray}
In order to obtain the best generalization 
ability,  we minimize the prefactor of 
$1/\sqrt{\alpha}$ in the second term of Eq. (\ref{type1}) 
and obtain 
\begin{equation}
{\eta}=\sqrt{\frac{\pi}{2}}
\frac{\Delta}
{a}.
\label{eta2}
\end{equation}
For this $\eta$, the condition 
$4a{\Delta}\eta/\sqrt{2\pi}-1>0$ is satisfied. 
In general, if we take $\eta$ independent of $a$, 
the condition $4a{\Delta}\eta/\sqrt{2\pi}-1>0$ is not always satisfied.
The quantity $b\,{\equiv}\,4a{\Delta}/\sqrt{2\pi}$ 
takes the maximum value $4/\sqrt{2\pi{e}}$ at $a=1$. 
Therefore, whatever value of $a$ we choose, we cannot obtain 
the ${\alpha}^{-1/2}$ convergence if 
the product of this maximum value $4/\sqrt{2\pi{e}}$ and 
$\eta$ is not larger than unity.
This means that 
$\eta$ should satisfy 
${\eta}>\sqrt{2{\pi}e}/4\,{\simeq}\,1.033$ 
for the first term of 
Eq. (\ref{sol}) dominate asymptotically, 
yielding Eq. (\ref{type1}), for 
a non-vanishing range of $a$. 
In contrast, if we choose $\eta$ to satisfy 
$b\,{\eta}-1<0$, 
the generalization error is dominated by the second term of 
Eq. (\ref{sol}) and 
behaves as 
\begin{equation}
{\epsilon}_{\rm g}=
2H(a)+\frac{\sqrt{2A}}{\pi}
\left(
\frac{\eta}{\alpha}
\right)^{2a\Delta\eta/\sqrt{2\pi}}.
\label{type2}
\end{equation}
In this case, the generalization 
error converges less quickly than (\ref{type1}). 
For example, if we choose ${\eta}=1$, 
we find that the condition 
$b\,{\eta} > 1$ cannot be satisfied by any $a$ 
and the generalization error converges 
as in Eq. (\ref{type2}). 
If we set $\eta=2$ ($> \sqrt{2\pi{e}}/4=1.033$) 
as another example, the asymptotic form of 
the generalization error 
is either Eq. (\ref{type1}) or Eq. (\ref{type2}) 
depending on the value of $a$. 
\section{CONCLUSION}
We have investigated the generalization abilities of 
a simple perceptron trained by  
the teacher who is also a simple perceptron but has a 
non-monotonic transfer 
function using the on-line AdaTron algorithm. 
For the learnable case ($a=\infty$), if we fix the 
length of the student weight vector as $l=|{\bf J}|/\sqrt{N}=1/2$, 
the generalization error 
converges to zero as ${\sim}\,3/(2\alpha)$ 
as Biehl and Riegler reported \cite{Biehl94}. 
However, if we allow the time 
development of the length of student 
weight vector, 
the asymptotic behavior of the 
generalization 
error shows dependence on the  
initial value of $l$. 
When the student starts the training process 
from the optimal length of weight vector $l$, 
we can obtain the generalization error  
${\epsilon}_{\rm g}\,{\sim}\,4/(3\alpha)$ which is 
a little faster than $3/(2\alpha)$. 
As the student is able to know the length of 
its own weight vector in principle, we can get the better 
generalization ability ${\epsilon}_{\rm g}\,{\sim}\,4/(3\alpha)$ 
by a heuristic search of the optimal initial value of 
$l$.
On the other hand, if the 
width $a$ of the reversed wedge has a finite value, 
the generalization error converges exponentially to a 
non-optimal $a$-dependent value. 
In addition, these residual errors are larger than 
those of the conventional perceptron learning for the 
whole range of $a$. 
Therefore  we conclude that, although the AdaTron learning 
is powerful for the learnable case \cite{Biehl94} including    
the situation in which the input vector is structured \cite{Riegler95}, 
it is not necessarily suitable for learning of 
the non-monotonic input-output relations.

Next we introduced the learning 
rate and 
optimized it. 
For the learnable case, the generalization error 
converges to zero 
as $\,{\sim}\,4/(3\alpha)$ which is 
as fast as the result obtained 
by selecting the optimal 
initial condition for the case of non-optimization, 
$g=1$. 
For this learnable case, the asymptotic form of 
the optimal leaning rate is $g_{\rm opt}\,{\sim}\,
3l/2$. 
Therefore, for the on-line AdaTron learning, 
it seems that the length of the student weight 
vector plays an important role  
to obtain a better generalization ability. 
If the task is unlearnable, 
the generalization error under optimized learning rate 
converges to the theoretical 
lower bound $2H(a)$ as 
${\sim}\,{\alpha}^{-1}$ for $a > a_{\rm c2}$. 
Using this strategy, we can get the optimal residual error 
for $a$ even exactly at $a_{\rm c1}$ for which the optimized perceptron 
learning failed to obtain the optimal residual error \cite{Inoue96}.

We also investigated 
the generalization ability 
using a parameter-free 
learning rate. 
When the task is learnable, we assumed $g_{\rm opt}={\eta}\,l$ 
and optimized the prefactor $\eta$. 
As a result, we obtained  ${\epsilon}_{\rm g}\,{\sim}\,
4/(3\alpha)$ which is the same 
asymptotic form as the 
parameter-dependent case. 
Therefore, we can obtain 
this generalization ability 
by a heuristic choice of $\eta$;  
we may choose the best $\eta$ 
by trial and error. 
On the other hand, 
for the unlearnable case, we used 
the asymptotic form 
of the $a$-dependent learning rate 
in the limit of ${\alpha}{\rightarrow}\infty$, 
$g_{\rm opt}\,{\sim}\,{\eta}\,l/{\alpha}$,  
and optimized the coefficient $\eta$.
The generalization error then 
converges to $2H(a)$ as ${\alpha}^{-1/2}$ 
for $b\,{\eta} > 1$. 
If  $b\,{\eta}<1$, 
 the generalization error 
decays to $2H(a)$ as 
${\alpha}^{-b\,{\eta}/2}$, where 
the exponent $b\,{\eta}/2$ is 
smaller than $1/2$ because $b\,{\eta}<1$. 
Similar slowing down of the convergence rate of the 
generalization error by 
tuning a control parameter was also reported by 
Kabashima and Shinomoto in the problem of 
learning of two-dimensional 
blurred dichotomy \cite{Kaba95}. 

In conclusion, we could overcome the difficulty of the 
AdaTron learning of unlearnable problems 
by optimizing the learning rate and the 
generalization 
error was shown to converge to the 
best possible value 
as long as the width $a$ of 
reversed wedge satisfies $a>a_{\rm c2}$. 
For the parameter region $a<a_{\rm c2}$, 
this approach does not work well 
because the optimal value 
of $R$ is $R_{*}$ instead of $1$; 
our optimization is designed 
to accelerate the increase 
to $R$ toward $1$.  

In this paper, we could construct  
a learning strategy suitable to achieve 
the $a$-dependent optimal value $2H(a)$  
for $a>a_{\rm c2}$. 
However, 
for $a<a_{\rm c2}$, it is a very 
difficult but challenging future 
problem to get the optimal value by improving  
the conventional AdaTron learning. 
\acknowledgements 
The authors would like to thank Dr. Yoshiyuki Kabashima 
for helpful suggestions and comments. 
One of the authors (J. I.) thanks Dr. Siegfried B$\ddot{\rm o}$s 
for several useful comments.

\begin{figure}
\caption{
Generalization error as a function of $R$ for 
$a=\infty$, $2$, $1$, $0.5$ and $a=0$.
}
\end{figure}
\begin{figure}
\caption{
Optimal overlap $R$ which gives the best possible 
value and overlaps which give the residual error 
for Hebbian, perceptron and AdaTron learning algorithms.
}
\end{figure} 
\begin{figure}
\caption{
Best possible value of the 
generalization error, 
the residual generalization errors of 
conventional Hebbian, perceptron  
and AdaTron learning algorithms are plotted as functions of $a$. 
Except for $a=\infty$ and $a=0$, the AdaTron 
learning cannot lead the student to the 
best possible value of the generalization error. 
In addition, for a finite value of $a$, the residual 
generalization error of the AdaTron learning is larger  
than that of the perceptron learning. 
}
\end{figure}
\begin{figure}
\caption{
$R$-$l$ trajectories of the AdaTron learning for the 
learnable case 
$a=\infty$. The fixed point depends on the initial value of 
$l=l_{\rm init}$. 
For the special case of $l_{\rm init}=0.5$, the flow of $l$ 
becomes independent of $\alpha$.
}
\end{figure}
\begin{figure}
\caption{
Generalization errors of the AdaTron, perceptron and Hebbian 
learning algorithms for the learnable case $a=\infty$. 
The initial value of $l$ is $l_{\rm init}=0.1$ for all algorithms.
The AdaTron learning shows the fastest convergence among the three.
}
\end{figure}

\begin{figure}
\caption{
$R$-$l$ trajectories of the AdaTron learning for the unlearnable case 
$a=2$. All flows of $l$ converge to the fixed 
point at $l_{0}=1/2$.
}
\end{figure}
\begin{figure}
\caption{
Generalization errors of the AdaTron, perceptron and Hebbian 
learning algorithms for the unlearnable case $a=2$. The 
AdaTron learning shows the largest residual error among the three.
}
\end{figure}
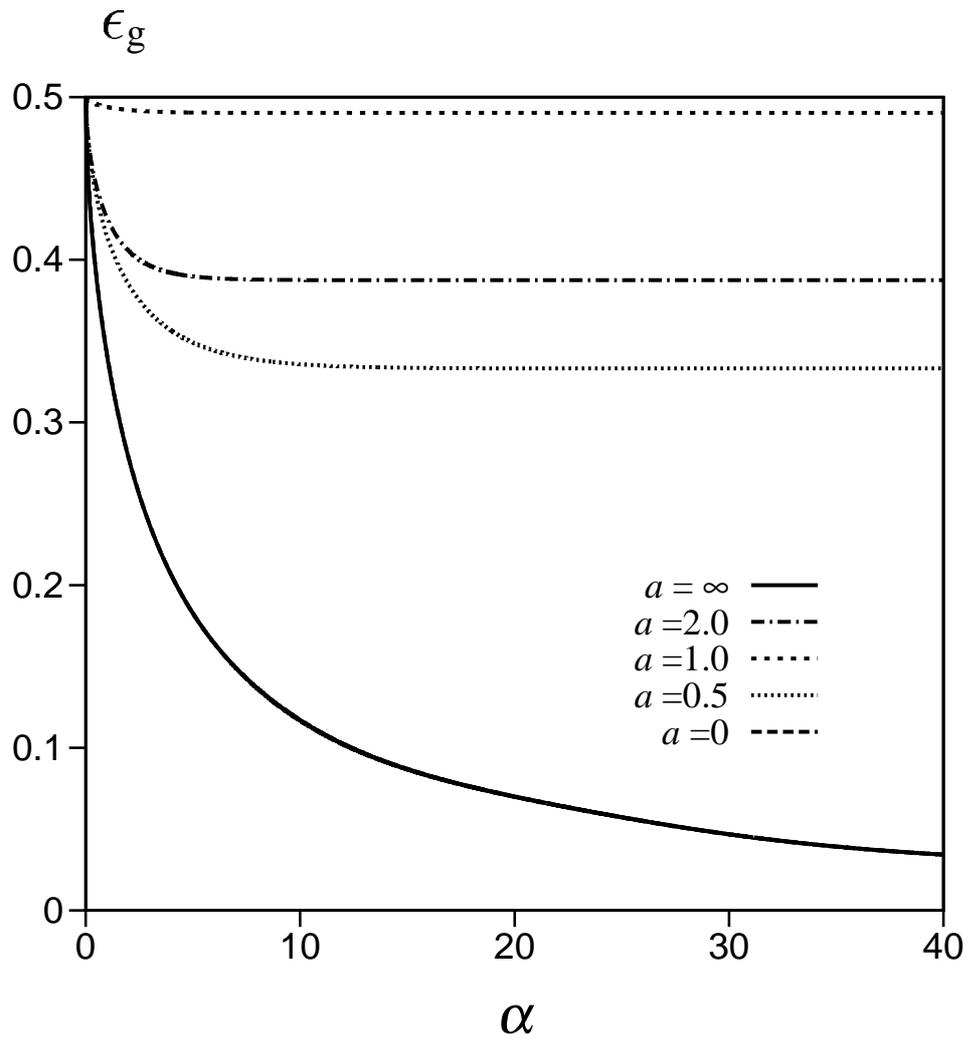
\begin{figure}
\caption{
Generalization errors of the AdaTron learning algorithm for the 
cases of $a=\infty$, $2$, $1$ and $0.5$. 
}
\end{figure}
\end{document}